\documentclass[a4paper,12pt]{article}
\usepackage[latin1]{inputenc}
\usepackage{graphicx}
\usepackage{amsmath,amssymb,euscript,array,mathrsfs}

\newcommand{\EQ}[1]{\begin{equation} #1 \end{equation}}
\newcommand{\AL}[1]{\begin{subequations} \begin{align} #1
\end{align}\end{subequations}}

\newcommand{\SP}[1]{\begin{equation}\begin{split} #1
\end{split}\end{equation}}

\begin{document}

\title{Glueball Spectra of SQCD-like Theories}
\author{
	\textbf{Daniel Elander}
	\bigskip
	\\
	{\it Department of Physics}, \\ {\it Swansea University}, \\ {\it Swansea, SA2 8PP, UK}
	\bigskip
	\\
	{\tt pyde@swansea.ac.uk}
}

\date{}

\maketitle

\abstract{We study the spectrum of scalar glueballs in SQCD-like theories whose gravity description is in terms of $N_c$ D5 color branes wrapped on an $S^2$ inside a CY3-fold, and $N_f$ backreacting D5 flavor branes wrapped on a non-compact two-cycle inside the same CY3-fold. We show that there exists a consistent truncation of the ten-dimensional Type IIB supergravity system to a five-dimensional non-linear sigma model consisting of four scalars coupled to gravity. Studying fluctuations of the scalars as well as the metric around particular backgrounds allows us to compute their spectra. A few different backgrounds share the same qualitative features, namely that the mass of the lightest scalar glueball increases as the number of flavors is increased, until one reaches the point $N_f = 2 N_c$ after which the opposite behaviour is obtained. We show that the five-dimensional non-linear sigma model obeys Seiberg duality, and demonstrate this explicitly for the spectra of a class of backgrounds that are Seiberg dual to themselves.}

\newpage
\tableofcontents
\newpage

\section{Introduction}

Since the AdS/CFT conjecture was originally proposed in \cite{Maldacena:1997re} and subsequently refined in \cite{Witten:1998qj, Gubser:1998bc}, it has been extended to apply to more generalized settings with less supersymmetry and also backgrounds that do not asymptote to AdS in the UV. In this paper, we will study the spectrum of scalar glueballs in SQCD-like theories, whose gravity description is in terms of $N_c$ D5 branes wrapping an $S^2$ inside a CY3-fold, and $N_f$ backreacting D5 flavor branes wrapping a non-compact two-cycle inside the same CY3-fold. The dual field theory is believed to be similar in the IR to $\mathcal N = 1$ SQCD with a quartic superpotential for the quark superfields \cite{Casero:2006pt}. However, the full theory cannot be dual to SQCD for a number of reasons. It does not have an $SU(N_f) \times SU(N_f) \times U(1)_R$ global symmetry as SQCD does, but instead only one $SU(N_f)$ (broken further to $U(1)^{N_f}$ by smearing the flavor branes as will be discussed later). Also, for $N_f < N_c$, the Affleck-Dine-Seiberg superpotential \cite{Affleck:1983mk} tells us that SQCD does not have a vacuum, whereas for the systems we will study backgrounds exist with $N_f < N_c$. The glueball spectra of the theories that we will study in this paper are not known since before. Using holographic techniques, we will find how the mass of the lightest scalar glueball in the spectrum depends on the number of flavors for a few different backgrounds.

The holographic prescription for computing the glueball spectrum is to study fluctuations around a particular background and look for solutions that satisfy correct boundary conditions in the IR and UV. This is only possible for specific values of $K^2 = - M^2$, where $K$ is the four-momentum of the fluctuations, and these special $K^2$ give us the spectrum. Technically, it is difficult to carry out such a study in ten dimensions. However, we will show that there exists a consistent truncation to a five-dimensional non-linear sigma model consisting of four scalars coupled to gravity. This will simplify the analysis. In \cite{Berg:2005pd}, an explicitly gauge-invariant formalism was developed for studying such systems, and formulas for the linearized equations of motion for the fluctuations were given in terms of a superpotential $W$. Here, we will derive the generalized versions of these formulas, which hold for an arbitrary potential $V$ not necessarily obtainable from a superpotential.

In the gravity picture, Seiberg duality is realized for these theories as a diffeomorphism, i.e. just a change of variables \cite{Casero:2006pt, HoyosBadajoz:2008fw}. Therefore, the background itself does not change under Seiberg duality, but since we have changed variables, the dictionary interpretation of the dual field theory is changed. We show that the Lagrangian of the five-dimensional non-linear sigma model is invariant under a set of transformations of the scalar fields and $N_c \rightarrow N_f - N_c$. It follows that anything that can be computed within this framework will obey Seiberg duality.

The backgrounds correponding to the setup described above that have been found fall into two categories known as Type A and Type N \cite{HoyosBadajoz:2008fw, Casero:2007jj}. Type A backgrounds are special cases of Type N backgrounds for which the VEV of the gaugino condensate as well as the mesons are zero. In this paper, we will study the spectrum of a few backgrounds of Type A for which the dilaton grows linearly in the UV. In the IR, there are different possible behaviours for the background (known as Type I, II and III \cite{Casero:2007jj}) corresponding to different vacua in the dual field theory. These backgrounds have a singularity in the IR which is ``good'' according to the criterion given in \cite{Maldacena:2000mw}, and are believed to capture the non-perturbative physics of the dual field theory. This criterion states that the $g_{00}$ component of the metric should not increase as we approach the singularity (the idea is that proper energy excitations should correspond to lower and lower energy excitations from the point of view of the field theory as one approaches the singularity in the IR).

The D5 flavor branes are smeared along the transverse angular coordinates, breaking the $SU(N_f)$ global symmetry to $U(1)^{N_f}$ (this procedure was first introduced in the context of flavor branes in \cite{Bigazzi:2005md}). The consistent truncation to five dimensions does not contain fluctuations of the gauge fields on the branes. However, it still contains fluctuations of the Ramond-Ramond 3-form $F_{(3)}$. Therefore, when $N_f \sim N_c$, the fluctuations that we consider mix glueballs and mesons. Since the fluctuations do not involve the gauge fields on the brane, the meson-glueballs whose spectrum we compute are $U(1)^{N_f}$-singlets.

Imposing the boundary condition on the fluctuations in the IR that their kinetic terms are regular, and in the UV that the fluctuations correspond to normalizable modes, we find that the mass of the lightest scalar glueball increases as the number of flavors is increased, until the point $N_f = 2 N_c$ is reached after which the opposite behaviour is observed. For a particular class of backgrounds that are Seiberg dual to themselves, we demonstrate explicitly that the spectrum obeys Seiberg duality.

There is by now a large literature on systems with back-reacting flavors. In the future, it would be interesting to apply the same techniques to study the glueball spectra of the various systems studied in \cite{HoyosBadajoz:2008fw, Casero:2007jj, Ramallo:2006et, Murthy:2006xt, Paredes:2006wb, Benini:2006hh, Casero:2007pz, Bertoldi:2007sf, Casero:2007ae, Benini:2007gx, Hirano:2007cj, Zeng:2007ta,  Burrington:2007qd, Benini:2007kg, Caceres:2007mu, Canoura:2008at, Cremonesi:2008zw, Bigazzi:2008gd, Bigazzi:2008zt, Bigazzi:2008ie, Bigazzi:2008cc, Arean:2008az, Gaillard:2008wt, Bigazzi:2008qq, Ramallo:2008ew, Bigazzi:2009gu, Caceres:2009bk, Bigazzi:2009bk, Nunez:2009da, Cotrone:2007qa}.

This paper is organized as follows. In section 2, we describe the general setup and the backgrounds that we will study. In section 3, we introduce the five-dimensional non-linear sigma model, discuss the Seiberg duality it obeys, and present the holographic techniques that we will use to compute the spectra. Section 4 contains the computation of the spectra. Finally, we summarize our results in section 5.

\section{Gravity Duals of SQCD-like Theories}

The backgrounds we will be interested in are obtained from wrapping $N_c$ D5 color branes on an $S^2$ inside a CY3-fold, then adding $N_f$ flavor branes that wrap a non-compact two-cycle inside the same CY3-fold. This is described in detail in \cite{Casero:2006pt}, where evidence is given for that the backgrounds obtained are dual to a field theory with similar behaviour in the IR as $\mathcal N = 1$ SQCD with a quartic superpotential for the quark superfields.

\subsection{Action and Equations of Motion}

We will now write the Type IIB supergravity action and the equations of motion that follow from it. The action (in Einstein frame) is given by
\SP{
	S = \hat S_{IIB}+ S^{(flavors)},
}
where $\hat S_{IIB}$ describes Type IIB supergravity in the truncation to the metric, the dilaton, and the RR 3-form $(g_{\mu\nu}, \phi, F_{(3)})$, and $S^{(flavors)}$ is the action of the flavor branes. We have that
\SP{
	\hat S_{IIB} = \frac{1}{2 \kappa_{(10)}^2} \int d^{10}
	x \sqrt{-g} \left[ R - \frac{1}{2} \partial_\mu \phi 
	\partial^\mu \phi - \frac{1}{12} e^{\phi} F_{(3)}^2 \right].
}
Choosing coordinates $(x^\mu, \rho, \theta, \varphi, \tilde \theta, \tilde \varphi, \psi)$, the flavor branes extend along the external coordinates $x^\mu$, the radial coordinate $\rho$, and the angular coordinate $\psi$. Their action is given by
\SP{
  S^{(flavors)} = T_{D5} \sum^{N_f} \Bigg[ - \int_{\mathcal{M}_6} d^6 x e^{\phi/2} \sqrt{- g_{(6)}} + \int_{\mathcal{M}_6} P[C_6] \Bigg],
}
where $g_{(6)}$ is the determinant of the pullback of the metric to $\mathcal{M}_6$, the world volume of the flavor brane, and similarly $P[C_6]$ is the pullback of $C_6$.
After smearing along the angular coordinates $(\theta, \varphi, \tilde \theta, \tilde \varphi)$, this becomes
\SP{
  S^{(flavors)} =
	\frac{T_{D5} N_f}{(4\pi)^2} \Bigg[ - \int 
	d^{10}x\sin\theta\sin\tilde{\theta} e^{\phi/2} \sqrt{-g_{(6)}} +
	\int C_6\wedge \Omega_4  \Bigg],
}
where
\SP{
	\Omega_4= \sin\theta \sin\tilde{\theta} 
	d\theta \wedge d\tilde{\theta} \wedge d\varphi\wedge d\tilde{\varphi}.
}

The equation of motion for the dilaton is
\SP{
	\frac{1}{\sqrt{-g}} \partial_\mu \left( g^{\mu \nu} \sqrt{-g} \partial_\nu \phi \right) - \frac{1}{12} e^\phi F_{(3)}^2 - \frac{N_f}{8} e^{\phi/2} \frac{\sqrt{-g_{(6)}}}{\sqrt{-g_{(10)}}} \sin \theta \sin \tilde \theta = 0,
}
while Einstein's equations read
\SP{
	R_{\mu\nu} - \frac{1}{2} g_{\mu\nu} R =& \frac{1}{2} \left( \partial_\mu \phi \partial_\nu \phi - \frac{1}{2} g_{\mu\nu} \partial_\lambda \phi \partial^\lambda \phi \right) + \\& \frac{e^\phi}{12} \left( 3 F_{\mu \sigma \lambda} F_\nu^{\ \sigma \lambda} - \frac{1}{2} g_{\mu\nu} F_{(3)}^2 \right) + T_{\mu\nu}^{(flavor)},
}
where
\SP{
	T^{\mu\nu}_{(flavor)} = - \frac{N_f}{8} \sin \theta \sin \tilde \theta e^{\phi/2} \delta^\mu_\alpha \delta^\mu_\beta g^{\alpha \beta}_{(6)} \frac{\sqrt{-g_{(6)}}}{\sqrt{-g_{(10)}}}.
}
Finally, Maxwell's equation for the $F_{(3)}$ is given by
\SP{
	\partial_\mu \left( \sqrt{-g} e^\phi F^{\mu\nu\lambda} \right) =0.
}

\subsection{Type A Backgrounds}
In this paper, we will be interested in so-called Type A backgrounds. For these backgrounds the VEV of the gaugino condensate is zero, as are the VEVs of the meson matrix. Type A backgrounds can be obtained from starting with the ansatz \cite{Casero:2006pt}
\SP{
	\label{eq:ansatzTypeA}
	ds^2 =& \mu^2 e^{2f} \Bigg[ \mu^{-2} dx_{1,3}^2 + e^{2k} d\rho^2 + e^{2h} (d\theta^2 + \sin^2 \theta d\varphi^2) + \\
	& \frac{e^{2\tilde g}}{4} (d\tilde\theta^2 + \sin^2 \tilde\theta d\tilde\varphi^2) + \frac{e^{2k}}{4} (d\psi + \cos \tilde\theta d\tilde\varphi + \cos \theta d\varphi)^2 \Bigg],
}
\SP{
	F_{(3)} = - \Big[ \frac{N_c}{4} \sin \tilde\theta d\tilde\theta \wedge d\tilde\varphi + \frac{N_f-N_c}{4} \sin \theta d\theta \wedge d\varphi \Big] \wedge &\\ (d\psi + \cos \tilde\theta d\tilde\varphi + \cos \theta d\varphi)&,
}
with $\mu^2 = \alpha' g_s$. Here, $f$, $k$, $h$, and $\tilde g$ are taken to be functions of the radial coordinate $\rho$. For the backgrounds that we will be interested in the IR is at $\rho = 0$ and the UV at $\rho = \infty$.

Making a change of variables to $P$, $Q$, and $Y$ through
\SP{
	e^{2h} =& \frac{P + Q}{4}, \\
	e^{2\tilde g} =& P - Q \\
	e^{2k} =& 4Y,
}
the BPS equations can be solved as ($f = \phi / 4$) \cite{HoyosBadajoz:2008fw, Casero:2007jj}
\SP{
	Q =& Q_0 + (2 N_c - N_f) \rho, \\
	Y =& \frac{1}{8} (P' + N_f), \\
	e^{4 (\phi - \phi_0)} =& \frac{e^{4\rho}}{(P^2 - Q^2) Y},
}
where $\phi_0$ and $Q_0$ are integration constants, and $P$ satisfies a second order differential equation given by
\SP{
	P'' + (P' + N_f) \left( \frac{P' + Q' + 2 N_f}{P - Q} + \frac{P' - Q' + 2N_f}{P + Q} - 4 \right) = 0.
}

\subsection{IR and UV Expansions}

\begin{figure}[p]
\centering
	\includegraphics[width=6cm]{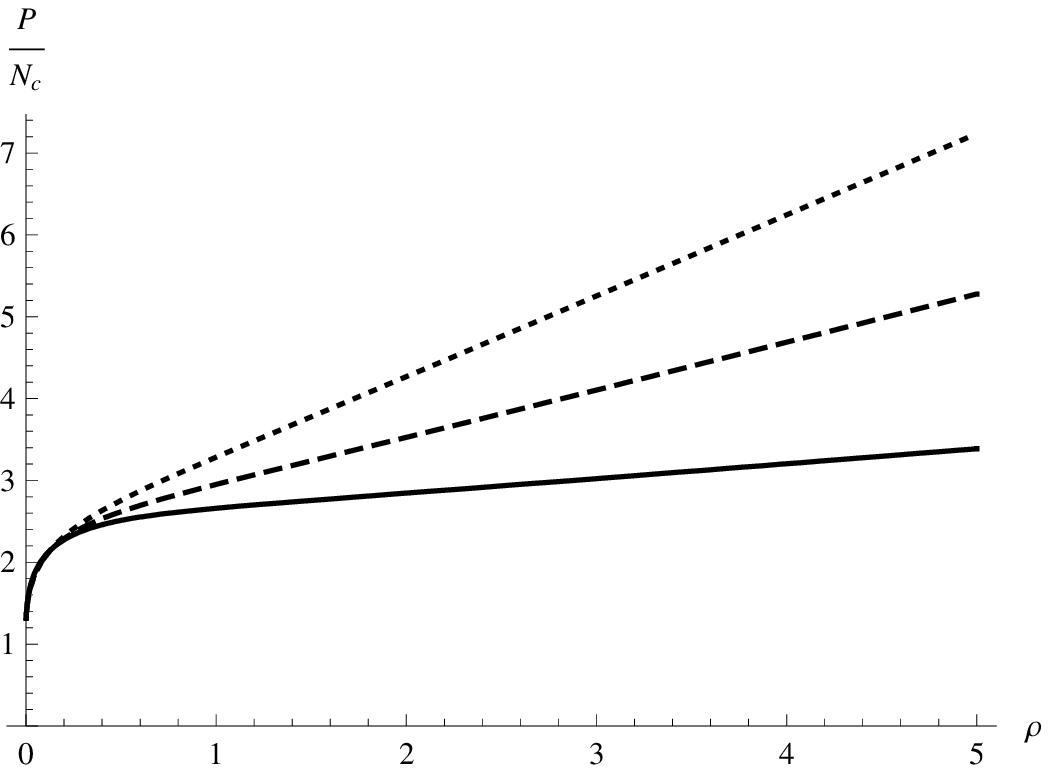}
	\caption{$P$ as a function of $\rho$ for the Type A background with Type II IR behaviour and $Q_0 = 1.2$. The different lines correspond to different number of flavors: dotted is $N_f = N_c$, dashed is $N_f = 1.4 N_c$, and solid is $N_f = 1.8 N_c$.}
	\label{fig:PplotTypeII_1}
\end{figure}

\begin{figure}[p]
\centering
	\includegraphics[width=6cm]{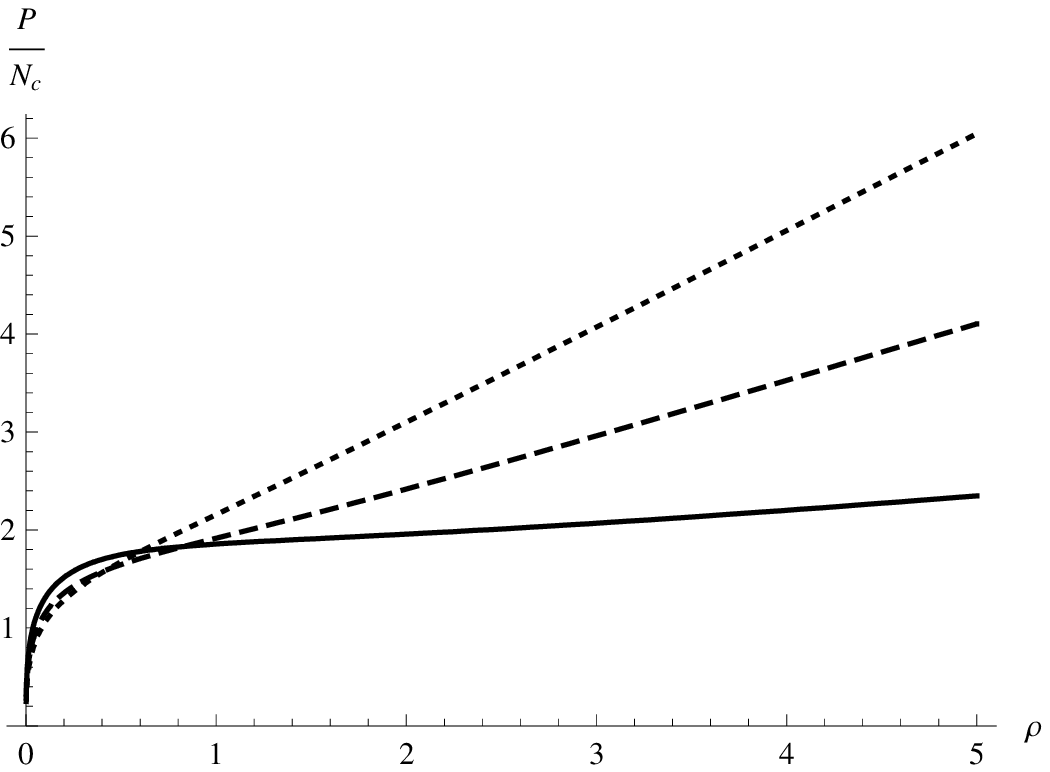}
	\caption{$P$ as a function of $\rho$ for the Type A background with Type III IR behaviour. The different lines correspond to different number of flavors: dotted is $N_f = N_c$, dashed is $N_f = 1.4 N_c$, and solid is $N_f = 1.8 N_c$.}
	\label{fig:PplotTypeIII_1}
\end{figure}

\begin{figure}[p]
\centering
	\includegraphics[width=6cm]{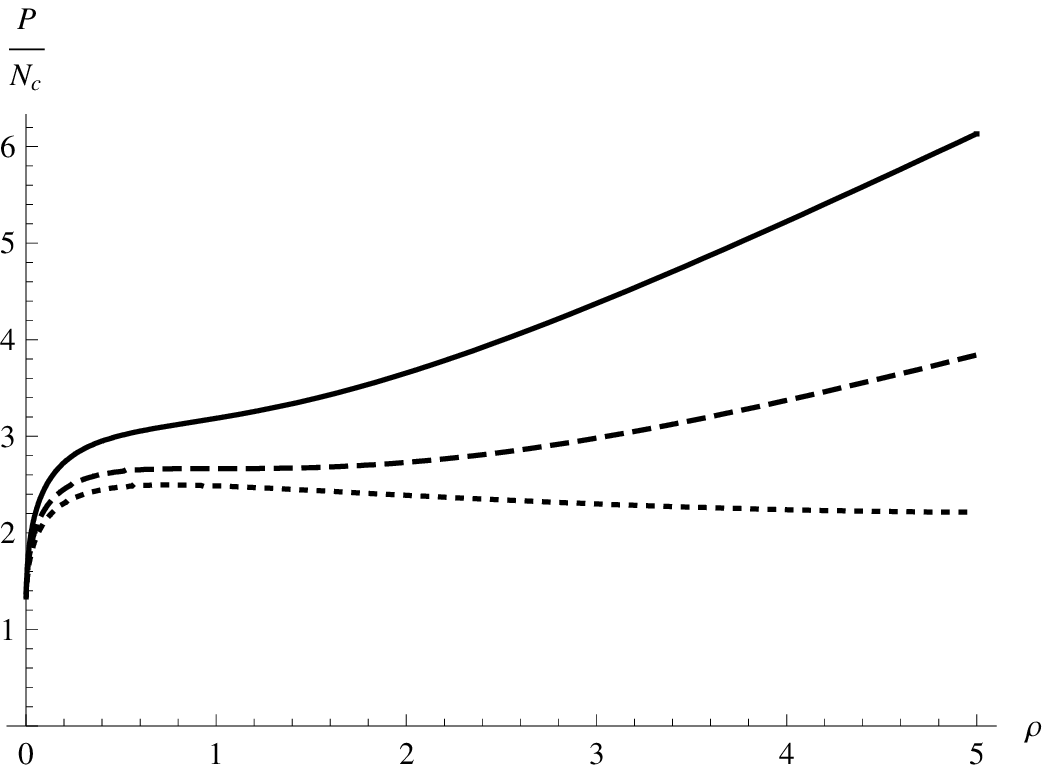}
	\caption{$P$ as a function of $\rho$ for the Type A background with Type II IR behaviour and $Q_0 = 1.2$. The different lines correspond to different number of flavors: dotted is $N_f = 2.2 N_c$, dashed is $N_f = 2.6 N_c$, and solid is $N_f = 3 N_c$.}
	\label{fig:PplotTypeII_2}
\end{figure}

\begin{figure}[t]
\centering
	\includegraphics[width=6cm]{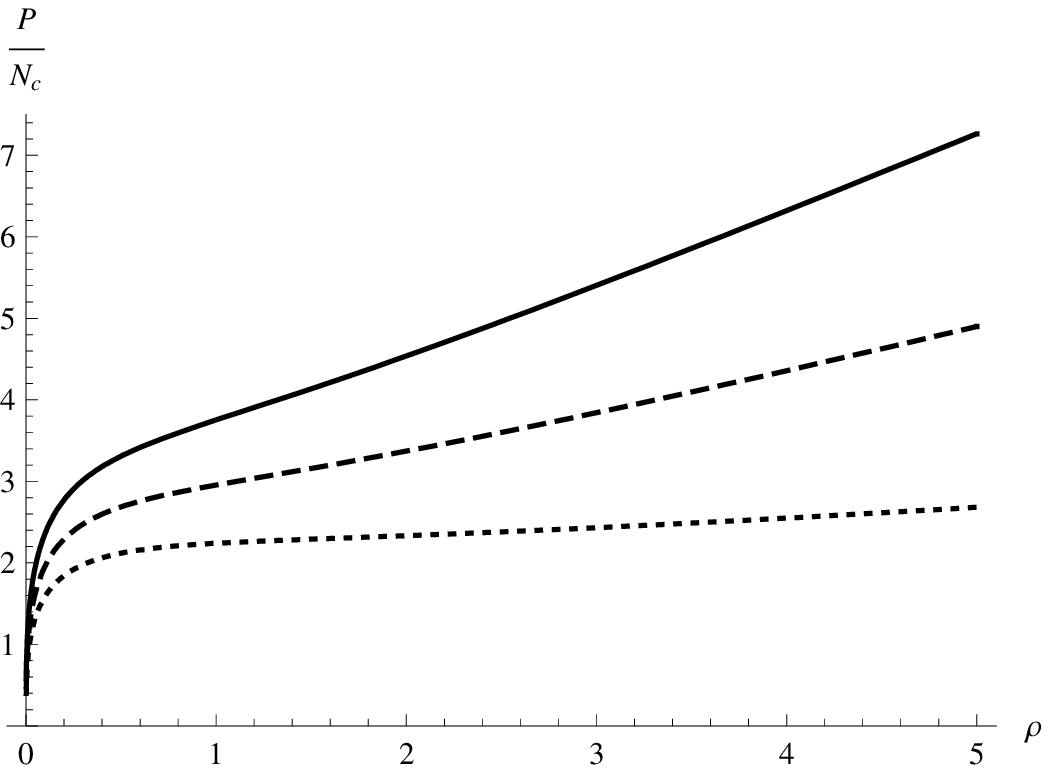}
	\caption{$P$ as a function of $\rho$ for the Type A background with Type III IR behaviour. The different lines correspond to different number of flavors: dotted is $N_f = 2.2 N_c$, dashed is $N_f = 2.6 N_c$, and solid is $N_f = 3 N_c$.}
	\label{fig:PplotTypeIII_2}
\end{figure}

We will be interested in backgrounds for which $P$ grows linearly in the UV.\footnote{The alternative is that the $P$ grows exponentially in the UV, in which case the spectrum only contains a continuum.} For $N_f < 2 N_c$, these have the UV expansion (around $\rho = \infty$) given by
\EQ{
\label{eq:UVexpansion}
	P_{UV} = (2 N_c - N_f) \rho + (1 + Q_0) N_c + \frac{N_f N_c}{4(2 N_c - N_f)} \rho^{-1} + \mathcal{O}\left(\rho^{-2}\right),
}
whereas if $N_f > 2 N_c$
\EQ{
	P_{UV} = - (2 N_c - N_f) \rho - ((1 + Q_0) N_c - N_f) - \frac{N_f (N_f - N_c)}{4(2 N_c - N_f)} \rho^{-1} + \mathcal{O}\left(\rho^{-2}\right).
}

In the IR, there are several different possible behaviours for the background. In this paper, we will focus on two different ones, that of Type II and Type III \cite{Casero:2007jj}. For Type II and $Q_0 > 0$ we have that \cite{Casero:2006pt}
\EQ{
	P_{IR}^{(II)} = Q_0 + 4 h_1 \sqrt{\rho} - \left( 2 N_c + \frac{8 h_1^2}{3 Q_0} + N_f \right) \rho + \mathcal{O}\left(\rho^{3/2}\right).
}
There are two integration constants: $Q_0$ and $h_1$. Solutions exist that interpolate smoothly between the Type II IR and the linear dilaton behaviour in the UV \cite{Casero:2006pt}. In order to obtain such solutions, one must dial the integration constants $Q_0$ and $h_1$, essentially making $h_1$ a function of $Q_0$. This leaves us with one free parameter $Q_0$ for the Type II solutions.

In the case of Type III, $Q_0 = 0$ and $P$ has the following behaviour in the IR:
\EQ{
	P_{IR}^{(III)} = 4 h_1 \rho^{1/3} - \frac{9 N_f}{5} \rho + \frac{8 h_1}{3} \rho^{4/3} + \mathcal{O}\left(\rho^{5/3}\right).
}
Now, requiring that the solution has the UV asymptotics of $P_{UV}$ completely fixes the one parameter $h_1$. Both the Type II and Type III backgrounds have a singularity in the IR, which satisfies the criterion for being a ``good'' singularity given in \cite{Maldacena:2000mw}.

Figure~\ref{fig:PplotTypeII_1} shows $P$ as a function of $\rho$ for the Type A background with Type II IR behaviour and $Q_0 = 1.2$ for a few different number of flavors $N_f < 2 N_c$. Figure~\ref{fig:PplotTypeIII_1} shows the corresponding plots for the Type A background with Type III IR behaviour. Figures~\ref{fig:PplotTypeII_2}~and~\ref{fig:PplotTypeIII_2} are the same as Figures~\ref{fig:PplotTypeII_1}~and~\ref{fig:PplotTypeIII_1} but for flavors $N_f > 2 N_c$.

\subsection{Dual Field Theory}

We will now describe in more detail some aspects of the field theory conjectured to be dual to the backgrounds studied above. First, consider the case of no flavors. Then the dual field theory is a four-dimensional $\mathcal N = 1$ supersymmetric field theory, obtained from a twisted compactification of six-dimensional SYM on $S^2$ where the twisting is such that it preserves four supercharges \cite{Andrews:2005cv, Andrews:2006aw}. At weak coupling, this field theory consists of a massless vector multiplet, $V$, as well as a Kaluza-Klein tower of massive chiral and vector multiplets, $\Phi_k$ and $V_k$. The infinite number of KK modes reflect the fact that the UV completion is not given by a quantum field theory (in fact, it is given by a Little String Theory). If it were possible to separate the scale set by the size of the $S^2$ from the scale $\Lambda$ at which the theory becomes strongly coupled, we would have a gravity dual of $\mathcal N = 1$ SYM. Unfortunately, this is not the case. The Lagrangian of the field theory without flavors has the generic form
\SP{
	\mathcal L =& \int d^4\theta \sum_k \left( \Phi^\dagger_k e^V \Phi_k + m_k \left| V_k \right|^2 \right) + \\&
	\int d^2\theta \left[ \mathcal W_\alpha \mathcal W^\alpha + \sum_k \left( W_{k,\alpha} W_k^\alpha + \mu_k \left| \Phi_k \right|^2 + \mathcal W (\Phi_k, V_k) \right) \right],
}
where $\mathcal W_a$ and $W_{k,\alpha}$ are the curvatures of $V$ and $V_k$, and $m_k$ and $\mu_k$ are the masses of the massive vector and chiral multiplets comprising the Kaluza-Klein tower. The superpotential $\mathcal W(\Phi_k, V_k)$, governing the interactions between the KK chiral and vector multiplets, is given by
\SP{
	\mathcal W(\Phi_k, V_k) = \sum_{i,j,k} z_{ijk} \Phi_i \Phi_j \Phi_k + \sum_k \hat f(\Phi_k) W_{k,\alpha} W_k^\alpha.
}

With the introduction of flavors, we also have to add the terms \cite{Casero:2006pt}
\SP{
	\int d^4 \theta \left( \bar Q^\dagger e^V \bar Q + Q^\dagger e^{-V} Q \right) + \int d^2 \theta \sum_{p,i,j,a,b} \kappa^{ij}_p \bar Q^{a,i} \Phi^{ab}_p Q^{b,j}
}
to the Lagrangian. Here, $a,b = 1, \ldots, N_c$ are indices of the fundamental and anti-fundamental representations of $SU(N_c)$ and $i,j = 1, \ldots, N_f$ are indices of the fundamental and anti-fundamental representations of $SU(N_f)$. Since the smearing procedure described above breaks $SU(N_f)$ to $U(1)^{N_f}$, the $\kappa^{ij}_p$ must serve the role of breaking this symmetry in the field theory, however its exact form is not known. In principle, we could have also considered the more general case where there is a superpotential for the flavors too. Again, the form of such a superpotential is not known.

As mentioned above, it is not possible to separate the scale set by the size of the $S^2$ from the scale $\Lambda$ at which the field theory becomes strongly coupled. Nevertheless, we can imagine integrating out at least some of the KK modes, if not all the way down to $\Lambda$. This then gives rise to an effective superpotential $W_{eff}$ containing quartic terms
\SP{
	W_{eff} \sim \sum_{p,i,j} \frac{\kappa^2_{p,i,j}}{2 \mu^2_p} (\bar Q_i Q_j)^2.
}
This means that in the IR the field theory is similar to $\mathcal N = 1$ SQCD with a quartic superpotential (bearing in mind that not all the KK modes can be integrated out in this fashion).

\section{5d Formalism}
\subsection{Non-linear Sigma Model}

There exists a consistent truncation of the 10d system to a 5d non-linear sigma model consisting of four scalars coupled to gravity. The action of this 5d model is obtained by plugging in the ansatz \eqref{eq:ansatzTypeA} into the Type IIB supergravity action and integrating over the angular coordinates (see Appendix A for a detailed derivation). After making a change of variables as
\SP{
	A &=\frac{1}{3} \left( 8f +2\tilde g +2h +k \right) - \log 2, \;\;\;
	g = -\tilde g + h + \log 2, \\
	p &= - \frac{1}{6} \left( 4 f + \tilde g + h +2k \right) + \frac{1}{2} \log 2, \;\;\;
	x = 2 f + \tilde g + h - \log 2,
}
and also changing to the radial coordinate as $dz = e^{A+k} d\rho$, one obtains ($\Phi = [g,p,x,\phi]$)
\SP{
	{\cal L}_{5d}^{eff}=\frac{4\mu^4 N_c^2 (4\pi)^3}{G_{10}} \sqrt{-g}\left[\frac{R}{4} - \frac{1}{2}G_{ab}\partial \Phi^a \partial \Phi^b - V(\Phi)\right],
}
where the non-linear sigma model metric $G_{ab}$ is given by
\SP{
	G_{ab} = \mathrm{diag} \left( \frac{1}{2},6,1,\frac{1}{4} \right),
}
and the potential $V$ is given by
\SP{
	V = \frac{1}{128} e^{-2 (g+2 (p+x))} \Big[ 16 \left(-4 e^{g+6 p+2 x}
   \left(1+e^{2 g}\right)+e^{4 g}+1\right)+ \\ 8 e^{2 g+6 p+x+\frac{\phi
   }{2}} N_f+e^{12 p+2 x+\phi } \left((N_f-N_c)^2+e^{4 g} N_c^2\right)\Big].
}
The backgrounds we will be interested in are of the form
\EQ{
	ds^2 = dz^2 + e^{2A}dx_{1,3}^2,
}
where $A$ is a warp factor.

We note that $V$ can be written in terms of a superpotential $W$ as follows:
\SP{
	V = \frac{1}{2} W^a W_a - \frac{4}{3} W^2,
}
with
\SP{
	W = \frac{1}{16} e^{-g-2 (p+x)} \Big[ e^{6 p+x+\frac{\phi }{2}}
   \left(\left(-1+e^{2 g}\right) N_c+N_f\right)- \\ 4
   \left(1+e^{2 g}+2 e^{g+6 p+2 x}\right)\Big],
}
where $W_a \equiv \partial W / \partial \Phi^a$, and indices are lowered and raised using the non-linear sigma model metric $G_{ab}$ and its inverse. The first order equations of motion following from the superpotential are
\SP{
	A' &= - \frac{2 W}{3}, \\
	\Phi'^a &= W^a,
}
and agree with the BPS equations given in Appendix B of \cite{Casero:2006pt}. In Appendix D, we show how to derive the expression for $W$ from the BPS equations.

The equations of motion for the scalars following from the 5d action read \cite{Berg:2005pd}
\SP{
	\tilde \nabla^2 \Phi^a + \mathcal{G}^a_{\ bc}\tilde g^{\mu\nu} (\partial_\mu \Phi^b) (\partial_\nu \Phi^c) - V^a = 0,
}
whereas Einstein's field equations read
\SP{
	- \tilde R_{\mu\nu} + 2 G_{ab} (\partial_\mu \Phi^a) (\partial_\nu \Phi^b) + \frac{4}{3} \tilde g_{\mu\nu} V = 0.
}
Here the tilde is used to denote a five-dimensional quantity, and $\mathcal{G}^a_{\ bc}$ is the Christoffel symbol with respect to the non-linear sigma model metric. One can verify that every solution to these 5d equations of motion also solves the full 10d Type IIB supergravity equations of motion for the particular choice of non-linear sigma model metric $G_{ab}$ and potential $V$ above. This shows that the five-dimensional non-linear sigma model is a consistent truncation of the 10d Type IIB supergravity system.

\subsection{Seiberg Duality}

For the models considered in this paper, Seiberg duality is realized on the gravity side as a diffeomorphism. While the background does not change, the change of variables means that the dictionary describing quantities in the dual quantum field theory changes.

In terms of the variables ($P$, $Q$, $Y$, $\phi$), Seiberg duality transforms \cite{Casero:2006pt}
\SP{
	Q &\rightarrow -Q, \\
	N_c &\rightarrow N_f - N_c,
}
leaving $P$, $Y$, and $\phi$ unchanged. Using the relations
\SP{
	e^{3A} =& \frac{e^{2 \phi} (P^2 - Q^2) \sqrt{Y}}{16}, \\
	e^{2g} =& \frac{P + Q}{P - Q}, \\
	e^{6p} =& \frac{4 e^{- \phi}}{\sqrt{P^2 - Q^2} Y}, \\
	e^{2x} =& \frac{e^{\phi} (P^2 - Q^2)}{16}
}
we see that in terms of the 5d variables, a Seiberg duality simply takes the form $g \rightarrow -g$. It is straightforward to see that both the potential $V$ and the non-linear sigma model metric $G_{ij}$ are invariant under this transformation. It follows that the whole 5d theory exhibits Seiberg duality, and therefore anything that we can compute within this framework, including the spectrum, will manifest Seiberg duality.

Considering that Seiberg duality is normally only a duality in the IR, it may seem odd that the whole 5d Lagrangian is invariant under the Seiberg duality transformations. However, in \cite{Strassler:2005qs} it is argued that $\mathcal N = 1$ SQCD with a quartic superpotential for the quark superfields can satisfy an {\it exact} Seiberg duality, where not only the IR of two different field theories are the same, but they in fact represent two different descriptions of the {\it same} renormalization group flow. The fact that in our setup Seiberg duality corresponds to diffeomorphisms supports the view that Seiberg duality is exact for these backgrounds. However, it is not clear how this happens from a field theory perspective. One can make the following schematic argument in the IR. Starting with a superpotential $W = h \mu^{-1} (\tilde Q Q)^2$, we first Seiberg dualize, obtaining $\hat W = h \mu M^2 + \tilde q M q$. For $N_f < 2 N_c$, $h$ is a relevant operator, so we can integrate out $M$, solving 0 = $\partial \hat W / \partial M = 2 h \mu M + \tilde q q$. Plugging $M$ back into $\hat W$ gives us an effective superpotential $\tilde W = - h^{-1} \mu^{-1} (\tilde q q)^2 / 2$. As can be seen, this superpotential is of the same form as the one that we started with, but describing a theory with $\tilde N_c = N_f - N_c$ colors and $\tilde N_f$ flavors. Furthermore, the coupling $h$ has been inverted, consistent with that for $\tilde N_f > \tilde 2 N_c$ the coupling appearing in front of the quartic term of the superpotential is irrelevant. It is, however, clear that this argument only works in the IR. In order to argue from the field theory that Seiberg duality is exact, one would have to take into account the KK modes that become important in the UV. From this point of view, it still remains somewhat mysterious why the backgrounds we are considering seem to have an exact Seiberg duality and what it means in terms of the dual field theory.

\subsection{Holographic Techniques for Computing \\ Spectra}

The holographic prescription for computing the spectrum of glueballs is to study flucuations around the background, expanding the equations of motion to linear order and look for solutions for the fluctuations that satisfy the correct boundary conditions in the IR and the UV. This is only possible for special values of $K^2$, where $K$ is the space-time momentum of a plane wave. These $K^2$ correspond to poles of the correlator $\left\langle \mathcal{O} \mathcal{O} \right\rangle$ (where $\mathcal O$ is the operator in the dual field theory corresponding to the fluctuation in question), and give us the scalar glueball spectrum of the dual field theory. In order to find them, we will employ a generalization of the numerical method described in \cite{Berg:2006xy} that in effect evolves solutions from both the IR and the UV, then determines whether they can be matched up smoothly at a midpoint (see Appendix E for a detailed description of the numerical methods used).

In \cite{Berg:2005pd}, an explicitly gauge invariant formalism was developed to compute spectra holographically. Formulas for the linearized equations of motions for the fluctuations were given in terms of a superpotential $W$. Here, we will give the generalized corresponding formulas that are valid for any potential $V$. Following \cite{Berg:2005pd}, we expand the metric and the scalar fields in fluctuations that are gauge invariant. First, we write the 5d metric in terms of a lapse function and shift vector $n^i$ as
\EQ{
	ds^2 = (n^i n_i + n^2) dz^2 + 2 n_i dx^i dz + g_{ij} dx^i dx^j.
}
Then, we expand to linear order in the fluctuations $(\mathfrak{a}^a,\mathfrak{b},\mathfrak{c},\mathfrak{d}^i,\mathfrak{e}_{ij})$ using the prescription given in \cite{Berg:2005pd}:
\SP{
	\Phi^a &\rightarrow \Phi^a + \mathfrak{a}^a, \\
	n &\rightarrow 1 + \mathfrak{b}, \\
	n^i &\rightarrow e^{2A} (\mathfrak{d}^i + \frac{\partial^i}{\Box} \mathfrak{c}), \\
	g_{ij} &\rightarrow e^{2A} (\eta_{ij} + \mathfrak{e}_{ij}).
}
Note that this is not just a probe calculation (in which the fluctuations of the metric are put to zero), but both the scalar fields and the metric are dynamical.

For a general potential $V$, the linearized scalar equations of motion are
\SP{
\label{eq:linearizedScalar}
	\Big[ D_z^2 + 4 A' D_z + e^{-2A} \Box \Big] \mathfrak{a}^a - (V^a_{\ |c} - \mathcal{R}^a_{\ bcd} \Phi'^b \Phi'^d) \mathfrak{a}^c - &\\
	\Phi'^a (\mathfrak{c} + \partial_z \mathfrak{b}) - 2 V^a \mathfrak{b} = &0.
}
Here, $V_a = \partial V / \partial \Phi^a$, and, as before, indices are lowered and raised by the sigma model metric $G_{ab}$ and its inverse. If an index is placed after a ``$|$'', it means that the covariant derivative with respect to $G_{ab}$ should be taken:
\EQ{
		A_{a|b} \equiv D_b A_a = \partial_b A_a - \mathcal{G}^c_{\ ab} A_c,
}
where
\EQ{
		\mathcal{G}^a_{\ bc} = \frac{1}{2} G^{ad} ( \partial_c G_{db} + \partial_b G_{dc} - \partial_d G_{bc} ).
}
The Riemann tensor with respect to the non-linear sigma model metric is given by
\EQ{
	\mathcal R^a_{\ bcd} = \partial_c \mathcal G^a_{\ bd} - \partial_d \mathcal G^a_{\ bc} + \mathcal G^a_{\ ce} \mathcal G^e_{\ bd} - \mathcal G^a_{\ de} \mathcal G^e_{\ bc}.
}
Also, a ``background-covariant'' derivative $D_z$ has been defined through
\EQ{
		D_z \varphi^a = \partial_z \varphi^a + \mathcal{G}^a_{\ bc} \Phi'^b \varphi^c.
}
Similarly, from the linearized Einstein's equations, we obtain
\SP{
\label{eq:linearizedEinstein}
	6 A'  \mathfrak{c} + 4 \Phi'_a (D_z \mathfrak{a}^a) - 4 V_a \mathfrak{a}^a - 8 V \mathfrak{b} &= 0, \\
	-\frac{1}{2} \Box \mathfrak{d}_i + 3 A' \partial_i \mathfrak{b} - 2 \Phi'_a \partial_i \mathfrak{a}^a &= 0,
}
where prime denotes the derivative with respect to the radial coordinate $z$, and $\Phi'_a \equiv G_{ab} \Phi'^b$.

As in \cite{Berg:2005pd}, \eqref{eq:linearizedEinstein} can be solved algebraically in momentum space giving the metric fluctuations $\mathfrak{b}$, $\mathfrak{c}$, and $\mathfrak{d}^i$ in terms of the scalar fluctuations $\mathfrak{a}^a$. Plugging back into \eqref{eq:linearizedScalar} then gives us a linear differential equation for just the scalars:
\SP{
\label{eq:flucdiff}
	&\Big[ D_z^2 + 4 A' D_z -e^{-2A} K^2 \Big] \mathfrak{a}^a - \\ &\Big[ V^a_{\ |c} - \mathcal{R}^a_{\ bcd} \Phi'^b \Phi'^d + \frac{4 
(\Phi'^a V_c + V^a \Phi'_c )}{3 A'} + \frac{16 V \Phi'^a \Phi'_c}{9 A'^2} \Big] \mathfrak{a}^c = 0.
}
In the special case where $V$ can be written in terms of a superpotential $W$, this formula agrees with the one given in \cite{Berg:2005pd}:
\SP{
\label{eq:diffeqscalars}
		\Bigg[ \left( \delta^a_b D_z + W^a_{|b} - \frac{W^a W_b}{W} - \frac{8}{3} W \delta^a_b \right) \left( \delta^b_c D_z - W^b_{|c} + \frac{W^b W_c}{W} \right) - \\ \delta^a_c e^{-2A} K^2 \Bigg] \mathfrak{a}^c = 0.
}

Changing the radial coordinate as $dz = e^{A+k} d\rho$, \eqref{eq:flucdiff} becomes
\SP{
\label{eq:flucdiffrho}
	\Big[ \delta^a_b \partial_\rho^2 + S^a_b \partial_\rho + T^a_b - \delta^a_b e^{2k} K^2 \Big] \mathfrak{a}^b = 0,
}
with
\SP{
	S^a_b =& 2 \mathcal{G}^a_{\ bc} \partial_\rho \Phi^c + 4 \left(\partial_\rho p + \partial_\rho A \right) \delta^a_b, \\
	T^a_b =& \partial_b \mathcal{G}^a_{\ cd} \partial_\rho \Phi^c \partial_\rho \Phi^d - \\& 4 e^{-8p} \Bigg[ \left( \frac{4 (V^a \partial_\rho \Phi^c + V^c \partial_\rho \Phi^a)}{3 \partial_\rho A} + \frac{16 V \partial_\rho \Phi^a \partial_\rho \Phi^c}{9 (\partial_\rho A)^2} \right) G_{cb} + \partial_b V^a \Bigg].
}
This is the second order linear differential equation for the scalar fluctuations that we need to solve for different values of $K^2$ imposing the correct boundary behaviour in the IR and UV. In the IR, we will require that the kinetic terms for the fluctuations are regular. In the UV, we will require that the fluctuations are normalizable.

\section{Scalar Spectra}

In this section, we will study some different Type A backgrounds, and compute the mass of the lightest scalar glueball as a function of the number of flavors. First, we will work out what the boundary conditions for the fluctuations are in the IR and UV. Then we will solve \eqref{eq:flucdiffrho} numerically for different values of $K^2$ using the methods outlined in Appendix E. This gives us the spectrum. In the following, we will put $x_f = N_f/N_c$, and rescale $P \rightarrow N_c P$ and similarly for $Q$ and $Y$. All masses given are in units of $\alpha' g_s N_c$.

\subsection{Boundary Conditions in the UV}
\subsubsection{$N_f < 2 N_c$}
We will now expand the differential equations for the scalars \eqref{eq:flucdiffrho} in the UV. For $N_f < 2 N_c$, the background is given by \eqref{eq:UVexpansion}. We obtain that
\SP{
	S^a_b =& 2 \delta^a_b + \mathcal{O}\left(\rho^{-1}\right), \\
	T =&
\left(
\begin{array}{llll}
 -4 & 0 & 4 & -2 \\
 0 & -6 & -1 & -\frac{1}{2} \\
 2 & -6 & -3 & \frac{1}{2} \\
 -4 & -12 & 2 & -3
\end{array}
\right) + \mathcal{O}\left(\rho^{-1}\right).
}
A basis that diagonalizes $T$ to leading order is given by
\SP{
	B =
\left(
\begin{array}{llll}
 -1 & \frac{2}{3} & 0 & 1 \\
 \frac{1}{2} & \frac{1}{6} & 1 & 0 \\
 1 & -\frac{1}{6} & -3 & \frac{1}{2} \\
 0 & 1 & -6 & -1
\end{array}
\right),
}
such that $B^{-1} T B$ is diagonal. To leading order, this diagonalizes the differential equations for the fluctuations, and we obtain four independent differential equations
\SP{
	\partial_\rho^2 \mathfrak{a}^1 + 2 \partial_\rho \mathfrak{a}^1 - (8 + K^2) \mathfrak{a}^1 =& 0 \\
	\partial_\rho^2 \mathfrak{a}^2 + 2 \partial_\rho \mathfrak{a}^2 - (8 + K^2) \mathfrak{a}^2 =& 0 \\
	\partial_\rho^2 \mathfrak{a}^3 + 2 \partial_\rho \mathfrak{a}^3 - K^2 \mathfrak{a}^3 =& 0 \\
	\partial_\rho^2 \mathfrak{a}^4 + 2 \partial_\rho \mathfrak{a}^4 - K^2 \mathfrak{a}^4 =& 0
}
with solutions $\mathfrak{a}^{1,2} \sim e^{(-1 \pm \sqrt{9 + K^2}) \rho}$ and $\mathfrak{a}^{3,4} \sim e^{(-1 \pm \sqrt{1 + K^2}) \rho}$. Note that if we imagine expanding the fluctuations as $\mathfrak{a}^a = \lambda^a(\rho) e^{ \left(-1 \pm \sqrt{C^a - M^2} \right) \rho}$ with $\lambda^a(\rho) = \sum_n \mathfrak{a}^a_{n} \rho^{b_{a,n}}$, the above analysis captures the $C^a$ but not the function $\lambda^a(\rho)$. Therefore, the exponential factors are in general multiplied by powers of $\rho$.\footnote{The validity of the expansion in powers of $\rho^{-1}$ hinges on that it is possible to find a basis in which the components with different exponential behaviour do not mix. In the case of \cite{Elander:2009pk}, this can be checked explicitly. However, in that case, $P$ is exponentially close to $P = 2 N_c \rho$ in the UV, so that we can always work with analytical expressions. In the present case, only the UV expansion of $P$ in powers of $\rho^{-1}$ is known. While we cannot verify that there exists a basis in which the different exponential behaviours do not mix, the fact that we can go to reasonably high cut-offs in the UV (around $\rho = 15$) before the numerics break down suggests that such a basis exists.} However, the exponential behaviour is all we need for setting up the boundary conditions in the numerics. We are interested in the subleading behaviour so we pick the minus signs. In \cite{Berg:2006xy}, a normalizability condition for the fluctuations was given:
\EQ{
		\int dz e^{2A} G_{ab} \psi^a \psi^b = \int d\rho e^{3A + k} G_{ab} \psi^a \psi^b < \infty.
}
In our case, we have in the UV that ($x_f = N_f / N_c$)
\EQ{
		e^{3A + k} = e^{2\rho + 2 \phi_0} \left[ \frac{\sqrt{1 - x_f}}{8} \rho^{1/2} + \mathcal{O}(\rho^{-1/2}) \right],
}
so that the subdominant fluctuations are always normalizable, while the dominant ones are not. Let us also point out that for $M^2 > 1$ or $M^2 > 9$, we start getting oscillatory behaviour for fluctuations in the UV, signalling the start of a continuum.

\subsubsection{$N_f > 2 N_c$}

Similar considerations as in the last section (but with a different $B$) now lead to solutions of the form $\mathfrak{a}^{1,2} \sim e^{(-1 \pm \sqrt{9 + (x_f - 1) K^2}) \rho}$ and $\mathfrak{a}^{3,4} \sim e^{(-1 \pm \sqrt{1 + (x_f - 1) K^2}) \rho}$ (times powers of $\rho$). Note that the appearance of $x_f$ is Seiberg duality at work (restoring units, it takes $g_s \alpha' N_c K^2 \rightarrow g_s \alpha' N_c (x_f - 1) K^2$).

\subsection{Boundary Conditions for Type II in the IR}
For Type II backgrounds, it is natural to expand the fluctuations in the IR as
\EQ{
\label{eq:flucansatzTypeII}
	\mathfrak{a}^a = \sum_{n=0}^\infty \mathfrak{a}^a_n \rho^{n/2}.
}
We will choose boundary conditions such that the kinetic terms of the scalars do not blow up, i.e. that the derivative $\partial_\rho \mathfrak{a}^a$ does not blow up in the IR. This fixes $\mathfrak{a}^a_1 = 0$ and, after plugging in the ansatz \eqref{eq:flucansatzTypeII} into the differential equations for the fluctuations \eqref{eq:flucdiffrho}, leads to four linearly independent solutions
\SP{
	\mathfrak{a}_{(1)} =&
	\begin{pmatrix} 1 \\ 0 \\ 0 \\ 0 \end{pmatrix} + 
	\begin{pmatrix} -2 +\frac{2 x_f}{Q_0} + \frac{16 h_1^2}{Q_0^2} \\ \frac{2 + 2 Q_0 - x_f}{2 Q_0} \\ \frac{2 + 4 Q_0 - x_f}{2 Q_0} \\ \frac{2 - x_f}{Q_0} \end{pmatrix} \rho +
	\mathcal{O}(\rho^{3/2}), \\
	\mathfrak{a}_{(2)} =&
	\begin{pmatrix} 0 \\ 1 \\ 0 \\ 0 \end{pmatrix} + 
	\begin{pmatrix} 12 + \frac{12}{Q_0} - \frac{6 x_f}{Q_0} \\ -6 + \frac{3x_f}{2 Q_0} + \frac{3}{h_1^2} \\ -12 + \frac{3x_f}{2 Q_0} + \frac{3}{h_1^2} \\ \frac{3x_f}{Q_0} + \frac{6}{h_1^2} \end{pmatrix} \rho +
	\mathcal{O}(\rho^{3/2}), \\
	\mathfrak{a}_{(3)} =&
	\begin{pmatrix} 0 \\ 0 \\ 1 \\ 0 \end{pmatrix} + 
	\begin{pmatrix} 4 + \frac{2}{Q_0} - \frac{x_f}{Q_0} \\ -2 + \frac{x_f}{4 Q_0} + \frac{1}{2h_1^2} \\ -4 + \frac{x_f}{4 Q_0} + \frac{1}{2h_1^2} \\ \frac{x_f}{2 Q_0} + \frac{1}{h_1^2} \end{pmatrix} \rho +
	\mathcal{O}(\rho^{3/2}), \\
	\mathfrak{a}_{(4)} =&
	\begin{pmatrix} 0 \\ 0 \\ 0 \\ 1 \end{pmatrix} + 
	\begin{pmatrix} \frac{2-x_f}{2 Q_0} \\ \frac{x_f}{8 Q_0} + \frac{1}{4h_1^2} \\ \frac{x_f}{8 Q_0} + \frac{1}{4h_1^2} \\ \frac{x_f}{4 Q_0} + \frac{1}{2 h_1^2} \end{pmatrix} \rho +
	\mathcal{O}(\rho^{3/2}).
}
This fixes our boundary conditions in the IR.

\subsection{Boundary Conditions for Type III in the IR}
For Type III backgrounds it is natural to expand the fluctuations as
\EQ{
	\mathfrak{a}^a = \sum_{n=0}^\infty \mathfrak{a}^a_n \rho^{n/3}.
}
We see that the requirement that the derivatives of the fluctuations do not blow up in the IR now leads to $\mathfrak{a}^a_1 = \mathfrak{a}^a_2 = 0$, which is a stronger requirement than for Type II, and consequently leads to fewer than four allowed linearly independent solutions in the IR:
\SP{
	\mathfrak{a}_{(1)} =&
	\begin{pmatrix} 0 \\ -\frac{1}{6} \\ 1 \\ 0 \end{pmatrix} + 
	\begin{pmatrix} 0 \\ -1 \\ -2 \\ 0 \end{pmatrix} \rho +
	\begin{pmatrix} 0 \\ -\frac{h_1 K^2}{12} \\ \frac{h_1 K^2}{2} \\ 0 \end{pmatrix} \rho^{4/3} + 
	\mathcal{O}(\rho^{5/3}), \\
	\mathfrak{a}_{(2)} =&
	\begin{pmatrix} 0 \\ -\frac{1}{12} \\ 0 \\ 1 \end{pmatrix} + 
	\begin{pmatrix} 0 \\ \frac{1}{2} \\ 1 \\ 0 \end{pmatrix} \rho +
	\begin{pmatrix} 0 \\ -\frac{h_1 K^2}{24} \\ 0 \\ -\frac{h_1 K^2}{2} \end{pmatrix} \rho^{4/3} + 
	\mathcal{O}(\rho^{5/3}).
}

Now that we have fixed the boundary conditions, singling out a number of allowed linearly independent solutions in the IR and UV, the question becomes whether for a particular value of $K^2$ it is possible to find linear combinations of the allowed solutions in the IR which when evolved towards the UV can be written as linear combinations of the allowed solutions in the UV. The numerical methods used for determining this are outlined in Appendix E.

\subsection{Results}

Figure~\ref{fig:AbelianFlavorsTypeII} shows the mass squared of the lightest scalar glueball as a function of $x_f = N_f/N_c$ for a couple of Type A backgrounds with Type II IR behaviour. As can be seen, the mass increases with the number of flavors, until the special point $N_f = 2 N_c$ (where the theory has certain peculiar properties) where it reaches the start of the continuum, $M^2 = 1$, and after that decreases as a function of the number of flavors.

\begin{figure}[t]
\centering
	\includegraphics[width=6cm]{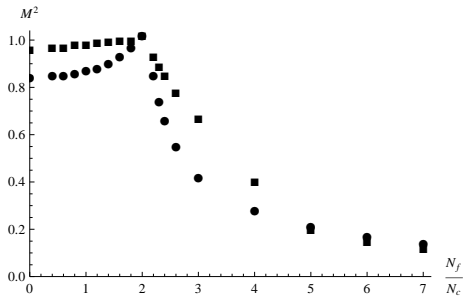}
	\caption{The mass squared of the lighest scalar glueball as a function of the number of flavors for a couple of Type A backgrounds with Type II IR behaviour: $Q_0=20$ (squares) and $Q_0=1.2$ (dots).}
	\label{fig:AbelianFlavorsTypeII}
\end{figure}

\begin{figure}[t]
\centering
	\includegraphics[width=6cm]{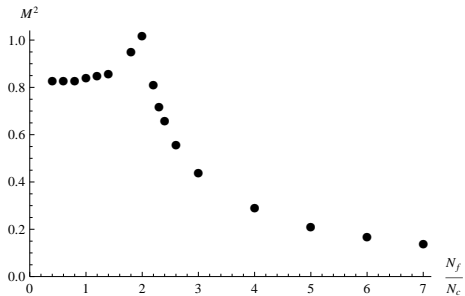}
	\caption{The mass squared of the lighest scalar glueball as a function of the number of flavors for the Type A background with Type III IR behaviour.}
	\label{fig:AbelianFlavorsTypeIII_1}
\end{figure}

Figure~\ref{fig:AbelianFlavorsTypeIII_1} shows the mass squared of the lighest scalar glueball as a function of $x_f = N_f / N_c$ for the Type A background with Type III IR behaviour. Again the same pattern can be seen.

\begin{figure}[t]
\centering
	\includegraphics[width=6cm]{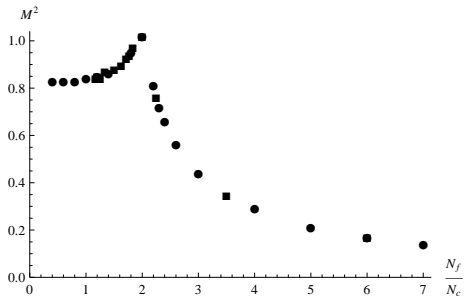}
	\caption{The mass squared of the lighest scalar glueball as a function of the number of flavors for the Type A background with Type III IR behaviour with the Seiberg dualized spectrum superimposed.}
	\label{fig:AbelianFlavorsTypeIII_2}
\end{figure}

Under Seiberg duality, the integration constant $Q_0 \rightarrow - Q_0$. Since for a Type A background with Type III IR behaviour $Q_0 = 0$, such backgrounds are Seiberg dual to themselves ($Q = (2 N_c - N_f) \rho \rightarrow (2 N_c - N_f) \rho$). In Figure~\ref{fig:AbelianFlavorsTypeIII_2}, the dots are the same as in Figure~\ref{fig:AbelianFlavorsTypeIII_1}, and the squares are what is obtained under Seiberg duality, mapping points as $x_f \rightarrow \frac{x_f}{x_f - 1}$. Also, under Seiberg duality, $M^2 \rightarrow (x_f - 1) M^2$, since we are working in units $\alpha' g_s N_c \rightarrow \alpha' g_s (N_f - N_c)$. As can be seen, the Seiberg dualized spectrum falls on the same trajectory.

\section{Conclusions}

We have been able to find a consistent truncation of the ten-dimensional Type IIB supergravity system describing $N_c$ D5 color branes and $N_f$ backreacting D5 flavor branes to five dimensions. The five-dimensional system is a non-linear sigma model coupled to gravity. In this model, Seiberg duality is realized at the level of the Lagrangian, i.e. any quantity that we can compute will automatically obey Seiberg duality.

We have computed the mass squared of the lighest scalar glueball for a few different Type A backgrounds, and found that the mass increases with the number of flavors for $N_f < 2 N_c$, but shows the opposite behaviour for $N_f > 2 N_c$. For a class of backgrounds that are Seiberg dual to themselves, we have seen explicitly how Seiberg duality is realized for the spectrum.

In the future, it would be interesting to apply the same techniques in order to compute the spectra of different systems with back-reacting flavors. For example, gravity duals that exhibit walking behaviour were found in \cite{Nunez:2008wi, Gurdogan:2009jd, Elander:2009pk}, and in particular one could imagine adding flavors to the walking backgrounds of \cite{Elander:2009pk} (for which $P$ grows linearly in the UV) and study how the spectrum is affected. It would be interesting to know what the effect of flavors is on the light scalar present for these backgrounds. We leave these questions for a future study.

\section*{Acknowledgments}

We would like to thank Maurizio Piai, Johannes Schmude, and especially Carlos N\'u\~nez for useful discussions and comments.

\appendix

\section{Deriving the 5d Effective Action}

Here, we will derive the 5d effective action of the non-linear sigma model for the more general case of Type N backgrounds, of which Type A is a special case. We start with the ansatz
\SP{
	\label{eq:ansatz}
	ds^2 =& \mu^2 e^{2f} \Bigg[ \mu^{-2} dx_{1,3}^2 + e^{2m} d\rho^2 + e^{2h} (d\theta^2 + \sin^2 \theta d\varphi^2) + \\
	& \frac{e^{2\tilde g}}{4} \left( (\tilde\omega_1 + a d\theta)^2 + (\tilde\omega_2 - a \sin \theta d\varphi)^2 \right) + \frac{e^{2k}}{4} (\tilde\omega_3 + \cos \theta d\varphi)^2 \Bigg], \\
	F_{(3)} =& \frac{\mu N_c}{4} \Bigg[-(\tilde\omega_1 + b d\theta) \wedge (\tilde\omega_2 - b \sin \theta d\varphi) \wedge (\tilde\omega_3 + \cos \theta d\varphi) + \\
	& dy^\mu \wedge (\partial_\mu b (-d\theta \wedge \tilde\omega_1 + \sin \theta d\varphi \wedge \tilde\omega_2)) + (1 - b^2) \sin \theta d\theta \wedge d\varphi \wedge \tilde\omega_3 \Bigg] - \\
	& \frac{\mu N_f}{4} \sin{\theta} d\theta \wedge d\varphi \wedge (d\psi +\cos{\tilde\theta} d\tilde\varphi),
}
where $\mu^2 = \alpha' g_s$, and
\SP{
	\tilde\omega_1 =& \cos \psi d\tilde\theta + \sin \psi \sin \tilde\theta d\tilde\varphi, \\
	\tilde\omega_2 =& -\sin \psi d\tilde\theta + \cos \psi \sin \tilde\theta d\tilde\varphi,\\
	\tilde\omega_3 =& d\psi + \cos \tilde\theta d\tilde\varphi.
}
First, let us consider the case when $a$, $b$, $f$, $\tilde g$, $h$, $k$, and $m$ only depend on $\rho$. Plugging \eqref{eq:ansatz} into the Type IIB action given by \cite{Casero:2006pt}
\AL{
	& & S_{10d} = \frac{1}{2 \kappa_{(10)}^2} \int d^{10}
	x \sqrt{-g} \left[ R - \frac{1}{2} \partial_\mu \phi 
	\partial^\mu \phi - \frac{1}{12} e^{\phi} F_{(3)}^2 \right] 
	- \\& & \frac{T_{D5} N_f}{(4\pi)^2}\Big(\int 
	d^{10}x\sin\theta\sin\tilde{\theta}\sqrt{g_{6}} - \int 
	C_6\wedge 
	\Omega_4  \Big), \nonumber\\
	& &  \Omega_4= \sin\theta \sin\tilde{\theta} 
	d\theta \wedge d\tilde{\theta} \wedge d\varphi\wedge d\tilde{\varphi},
\label{actionwithflavors}
}
and performing the integration over the angular coordinates yields
\EQ{
	S_{10d} = \frac{4 \mu^4 (4\pi)^3}{2 \kappa_{(10)}^2} \int d^4 x \int d\rho e^{8f+2\tilde g+2h+k+m} \left( T - V \right),
}
where
\SP{
	T = e^{-2m} \Bigg[& -\frac{e^{2 \tilde{g}-2 h}}{128} a'^2-\frac{N_c^2 e^{-4 f-2 h+\phi -2 \tilde{g}}}{128}
   b'^2+\frac{9f'^2}{4}+\frac{h'^2}{16}- \\& \frac{\phi'^2}{64}+\frac{\tilde{g}'^2}{16} +f'
   h'+\frac{f' k'}{2}+\frac{h' k'}{8}+f' \tilde{g}'+\frac{1}{4} h'
   \tilde{g}'+\frac{1}{8} k' \tilde{g}' \Bigg],
}
and
\SP{
	V =& \frac{e^{-2 \left(2 (f+h)+k+2 \tilde{g}\right)}}{256} \times \\
   \Bigg[ & 8 e^{2 \left(2 f+h+2 k+\tilde{g}\right)} a^2+8
   e^{4 f+2 h+6 \tilde{g}} a^2+ 16 e^{4
   (f+h+k)}+ \\& \left(a^2-1\right)^2 e^{4
   \left(f+k+\tilde{g}\right)}- 64 e^{2 \left(2
   (f+h)+k+\tilde{g}\right)}- \\& 16 \left(a^2+1\right) e^{2
   \left(2 f+h+k+2 \tilde{g}\right)}+ 16 e^{4 h+\phi }
   N_c^2+ 8 (a-b)^2 e^{2 h+\phi +2 \tilde{g}}
   N_c^2+ \\& e^{\phi +4 \tilde{g}}
   \left(N_f- \left(a^2-2 b a+1\right)
   N_c\right)^2+ 8 e^{\frac{1}{2} \left(4
   (f+h+k)+\phi +4 \tilde{g}\right)} N_f \Bigg].
}
Notice  that the Wess-Zumino term, whose only effect is to change the Bianchi identity of $F_3$, does not appear in this action, from which the Einstein, dilaton and Maxwell equations are derived.

Let us change coordinates to
\SP{
& f = A + p - \frac{x}{2}, \;\;
\tilde g = -A - \frac{g}{2} + \log 2  - p + x, \\
& h = -A + \frac{g}{2} - p + x, \;\;\; k = -A + \log 2  - 4 p, 
}
with inverse
\SP{
	A &=\frac{1}{3} \left( 8f +2\tilde g +2h +k \right) - \log 2, \;\;\;
	g = -\tilde g + h + \log 2, \\
	p &= - \frac{1}{6} \left( 4 f + \tilde g + h +2k \right) + \frac{1}{2} \log 2, \;\;\;
	x = 2 f + \tilde g + h - \log 2,
}
and also change the radial coordinate as $dz = e^{A+m} d\rho$. This leads to
\EQ{
	S_{10d} = \frac{4 \mu^4 N_c^2 (4\pi)^3}{2 \kappa_{(10)}^2} \int d^4 x \int dz e^{4A} \left( T - V \right),
}
with
\EQ{
	T = 3A'^2-\frac{1}{4} e^{-2 g} a'^2-\frac{N_c^2 e^{\phi -2 x}}{64} b'^2-\frac{g'^2}{4}-3p'^2-\frac{x'^2}{2}-\frac{\phi'^2}{8},
}
and
\SP{
	V =& \frac{e^{-2 (g+2 (p+x))}}{128} \times \\ \Bigg[& e^{12 p+2 x+\phi }
   \left(2 e^{2 g} (a-b)^2+e^{4 g}+\left(a^2-2 b
   a+1\right)^2\right) N_c^2- \\& 2 \left(a^2-2 b a+1\right)
   e^{12 p+2 x+\phi } N_f N_c+e^{12 p+2 x+\phi } N_f^2+ 8 e^{2 g+6
   p+x+\frac{\phi }{2}} N_f + \\& 16
   \left(a^4+2 \left(\left(e^g-e^{6 p+2
   x}\right)^2-1\right) a^2+e^{4 g}-4 e^{g+6 p+2 x}
   \left(1+e^{2 g}\right)+1\right) \Bigg]
}
Recognizing that for a metric given by $ds_5^2 = dz^2 + e^{2A} dx_{1,3}^2$ the Ricci scalar is (up to partial integrations) equal to $R = -12A'^2$, we can write this as the action of a 5d non-linear sigma model
\SP{
	S_{5d} = \frac{4 \mu^4 (4\pi)^3}{2 \kappa_{(10)}^2} \int d^4 x \int dz \sqrt{-g} \Bigg[ \frac{R}{4} - \frac{1}{2} G_{ij} \partial_\mu \Phi^i \partial^\mu \Phi^j - V(\vec \Phi) \Bigg],
}
where $\Phi = [g, p, x, \phi, a, b]$, and the non-linear sigma model metric is diagonal with entries $G_{gg} = \frac{1}{2}$, $G_{pp} = 6$, $G_{xx} = 1$, $G_{\phi \phi} = \frac{1}{4}$, $G_{aa} = \frac{e^{-2g}}{2}$, and $G_{bb} = \frac{N_c^2 e^{-2x+\phi}}{32}$.

\section{Seiberg Duality for Type N}

We can generalize the arguments of section 3.4 to Type N. For these backgrounds, it is convenient to go to the variables $P$, $Q$, $Y$, $\tau$, and $\sigma$ defined through \cite{HoyosBadajoz:2008fw}
\SP{
	e^{2h} =& \frac{1}{4} \left( \frac{P^2 - Q^2 }{P \cosh \tau - Q} \right), \\
	e^{2g} =& P \cosh \tau - Q , \\
	e^{2k} =& 4 Y, \\
	a =& \frac{P \sinh}{P \cosh \tau - Q}, \\
	b =& \frac{\sigma}{N_c}.
}
Seiberg duality corresponds to the transformation
\SP{
	Q &\rightarrow -Q, \\
	\sigma &\rightarrow -\sigma, \\
	N_c &\rightarrow N_f - N_c,
}
leaving $P$, $Y$, $\phi$, and $\tau$ unchanged.
Using the relations
\SP{
	e^{3A} =& \frac{e^{2 \phi} (P^2 - Q^2) \sqrt{Y}}{16}, \\
	e^{2g} =& \frac{P^2 - Q^2}{(\cosh \tau P- Q)^2}, \\
	e^{6p} =& \frac{4 e^{- \phi}}{\sqrt{P^2 - Q^2} Y}, \\
	e^{2x} =& \frac{e^{\phi} (P^2 - Q^2)}{16}
}
we see that in terms of the 5d variables, a Seiberg duality takes the form
\SP{
	e^g &\rightarrow \frac{e^g}{e^{2g} + a^2}, \\
	a &\rightarrow \frac{a}{e^{2g} + a^2}, \\
	b &\rightarrow \frac{N_c}{N_c-N_f} b, \\
	N_c &\rightarrow N_f - N_c.
}
Again, it is straightforward to see that both the non-linear sigma model metric $G_{ab}$ and the potential $V$ of the previous section are invariant under these transformations. It follows that the whole 5d theory obeys Seiberg duality.

\section{Linearized Equations of Motion}

Here we will generalize the results of \cite{Berg:2005pd} to cases where the potential $V$ can not necessarily be written in terms of a superpotential. The discussion closely follows that of \cite{Berg:2005pd}.

\subsection{ADM formalism}

We start by writing the metric on the form
\SP{
	\tilde g_{\mu\nu} =
	\left(
	\begin{array}{ll}
	 	n_i n^i + n^2 & n_j \\
	 	n_i & g_{ij}
	\end{array}
\right),
}
where the indices $i$ and $j$ run over the four-dimensional space-time, and there is also the radial coordinate (index $0$). The tilde is used to refer to five-dimensional quantities. The inverse metric is given by
\SP{
	\tilde g^{\mu\nu} =
	\frac{1}{n^2}
	\left(
	\begin{array}{ll}
	 	1 & -n^j \\
	 	-n^i & n^2 g^{ij} + n^i n^j
	\end{array}
\right).
}
The second fundamental form is
\SP{
	\mathcal{K}_{ij} = n \tilde \Gamma^0_{ij} = -\frac{1}{2n} (\partial_0 g_{ij} - \nabla_i n_j - \nabla_j n_i),
}
and one can derive the following relations
\SP{
\label{gammarelations}
	\tilde \Gamma^k_{ij} =& \Gamma^k_{ij} - \frac{n^k}{n} \mathcal{K}_{ij}, \\
	\tilde \Gamma^0_{i0} =& \frac{1}{n} \partial_i n + \frac{n^j}{n} \mathcal{K}_{ij}, \\
	\tilde \Gamma^k_{i0} =& \nabla_i n^k - \frac{n^k}{n} \partial_i n - n \mathcal{K}_{ij} \left( g^{jk} + \frac{n^j n^k}{n^2} \right), \\
	\tilde \Gamma^0_{00} =& \frac{1}{n} (\partial_0 n + n^j \partial_j n + n^i n^j \mathcal{K}_{ij}), \\
	\tilde \Gamma^k_{00} =& \partial_0 n^k + n^i \nabla_i n^k - n \nabla^k n - 2n \mathcal{K}^k_i n^i - n^k \tilde \Gamma^0_{00}.
}

\subsection{Expanding the Scalar EOMs}

The equation of motion for the scalars following from the 5d action is
\SP{
	\tilde \nabla^2 \Phi^a + \mathcal{G}^a_{bc}\tilde g^{\mu\nu} (\partial_\mu \Phi^b) (\partial_\nu \Phi^c) - V^a = 0.
}
Using the relations of the previous section to rewrite this in terms of 4d quantities, we obtain
\SP{
	\Big\{ \partial_z^2 - 2n^i \partial_i \partial_z + n^2 \nabla^2 + n^i n^j \nabla_i \partial_j - (n \mathcal{K}^i_i + \partial_z \ln n - n^i \partial_i \ln n) \partial_z + &\\
	\left[ n \nabla^i n - \partial_z n^i + n^j \nabla_j n^i + n^i (n \mathcal{K}^j_j + \partial_z \ln n - n^j \partial_j \ln n) \right] \partial_i \Big\} \Phi^a + &\\
	\mathcal{G}^a_{\ bc} \Big[ (\partial_z \Phi^b) (\partial_z \Phi^c) - 2n^i (\partial_i \Phi^b) (\partial_z \Phi^c) + &\\ (n^2 g^{ij} + n^i n^j) (\partial_i \Phi^b) (\partial_j \Phi^c) \Big] - n^2 G^{ab} \frac{\partial V }{\partial \Phi^b} = 0&.
}
We will now expand around a background (assumed to be dependent only on $z$) to linear order as follows:
\SP{
	\phi^a &\rightarrow \Phi^a + \mathfrak{a}^a, \\
	n &\rightarrow 1 + \mathfrak{b}, \\
	n^i &\rightarrow e^{2A} (\mathfrak{d}^i + \frac{\partial^i}{\Box} \mathfrak{c}), \\
	g_{ij} &\rightarrow e^{2A} (\eta_{ij} + \mathfrak{e}_{ij}).
}
It will be useful that
\SP{
	n \mathcal{K}^i_j \rightarrow - \partial_z A + \frac{1}{2} \left( \partial^i \mathfrak{d}_j + \partial_j \mathfrak{d}^i + 2 \frac{\partial^i \partial_j}{\Box} \mathfrak{c} - \partial_z \mathfrak{e}^i_j \right),
}
and
\SP{
	n \mathcal{K}^i_i \rightarrow - d \partial_z A + \mathfrak{c},
}
where $d + 1$ is the number of space-time dimensions (so that in our case $d = 4$).
At zeroth order, we have
\SP{
\label{eq:zerothorderscalar}
	\Phi''^a = - d A' \Phi'^a - \mathcal{G}^a_{\ bc} \phi'^b \Phi'^c + V^a,
}
whereas at first order, we obtain
\SP{
\label{eq:scalarfirstorder}
	&\partial_z^2 \mathfrak{a}^a + e^{-2A} \Box \mathfrak{a}^a + d A' \partial_z \mathfrak{a}^a + 2 \mathcal{G}^a_{\ bc} \Phi'^b \partial_z \mathfrak{a}^c + \\& \partial_d \mathcal{G}^a_{\ bc} \Phi'^b \Phi'^c a^d - \frac{\partial V^a}{\partial \Phi^c} \mathfrak{a}^c - \Phi'^a (\mathfrak{c} + \partial_z \mathfrak{b}) - 2 V^a \mathfrak{b} = 0.
}
Defining a ``background covariant'' derivative as
\SP{
	D_z \varphi^a = \partial_z + \mathcal{G}^a_{\ bc} \Phi'^b \varphi^c,
}
and using \eqref{eq:zerothorderscalar}, we can write \eqref{eq:scalarfirstorder} as
\SP{
\label{eq:scalarfirstorder2}
	\Big[ D_z^2 + d A' D_z + e^{-2A} \Box \Big] \mathfrak{a}^a - &(V^a_{|c} - \mathcal{R}^a_{\ bcd} \Phi'^b \Phi'^d) \mathfrak{a}^c - \\& \Phi'^a (\mathfrak{c} + \partial_z \mathfrak{b}) - 2 V^a \mathfrak{b} = 0.
}

\subsection{Expanding Einstein's Equations}

Einstein's equations are
\SP{
	E_{\mu\nu} = - \tilde R_{\mu\nu} + 2 G_{ab} (\partial_\mu \Phi^a) (\partial_\nu \Phi^b) + \frac{4}{d-1} \tilde g_{\mu\nu} V = 0.
}
In terms of four-dimensional quantities, this can be written as (normal, mixed, and tangential components)
\SP{
	(n \mathcal{K}^i_j) (n \mathcal{K}^j_i) - (n \mathcal{K}^i_i)^2 + n^2 R - 4n^2 V + 2G_{ab} \Big[ (\partial_z \Phi^a) (\partial_z \Phi^b) - &\\ 2n^i (\partial_i \Phi^a) (\partial_z \Phi^b) + (n^i n^j - n^2 g^{ij}) (\partial_i \Phi^a) (\partial_j \Phi^b) \Big] = 0&,
}
\SP{
	\partial_i (n \mathcal{K}^j_j) - \nabla_j (n \mathcal{K}^j_i) - n \mathcal{K}^j_j \partial_i \ln n + n \mathcal{K}^j_i \partial_j \ln n - &\\ 2G_{ab} (\partial_z \Phi^a - n^j \partial_j \Phi^a) \partial_i \Phi^b = 0&,
}
and
\SP{
	- \partial_z (n \mathcal{K}^i_j) + n^k \nabla_k (n \mathcal{K}^i_j) + n \mathcal{K}^i_j (n \mathcal{K}^k_k + \partial_z \ln n - n^k \partial_k \ln n) +  &\\ n \nabla^i \partial_j n +
	n \mathcal{K}^i_k \nabla_j n^k - n \mathcal{K}^k_j \nabla_k n^i - n^2 R^i_j + &\\ 2n^2 G_{ab} (\nabla^i \Phi^a) (\partial_j \Phi^b) + \frac{4 n^2 V}{d-1} \delta^i_j = 0&.
}

At zeroth order, the normal and tangential components together yield (written on a form suggestive of the superpotential formalism)
\SP{
\label{eq:Einsteinfirstorder}
	V =& \frac{1}{2} G_{ab} \Phi'^a \Phi'^b - \frac{d (d-1)}{4} A'^2, \\
	A'' =& - \frac{2}{d-1} G_{ab} \Phi'^a \Phi'^b.
}
At first order, the normal component gives
\SP{
\label{eq:normalfirstorder}
	2(d-1) A' \mathfrak{c} + 4 \Phi'_a (D_z \mathfrak{a}^a) - 4 V_a \mathfrak{a}^a - 8 V b = 0,
}
where we use the notation $\Phi'_a \equiv G_{ab} \Phi'^b$. The mixed components give
\SP{
\label{eq:mixedfirstorder}
	-\frac{1}{2} \Box \mathfrak{d}_i + (d-1) A' \partial_i \mathfrak{b} - 2 \Phi'_a \partial_i \mathfrak{a}^a = 0.
}
Here, we have used that $\mathfrak{e}^i_j$ and $\mathfrak{d}_i$ are transverse, and that $R^i_j = -\frac{1}{2} e^{-2A} \Box \mathfrak{e}^i_j$ and $R = 0$ at first order.

$\mathfrak{d}^i = 0$ to first order (decompose \eqref{eq:mixedfirstorder} as transverse and tangential), and therefore \eqref{eq:mixedfirstorder} implies that
\SP{
	\mathfrak{b} = \frac{2 \Phi'_a \mathfrak{a}^a}{(d-1) A'}.
}
Plugging into \eqref{eq:normalfirstorder} gives
\SP{
	\mathfrak{c} = \frac{8V \Phi'_a \mathfrak{a}^a}{(d-1)^2 A'^2} + \frac{2 V_a \mathfrak{a}^a}{(d-1) A'} - \frac{2 \Phi'_a D_z \mathfrak{a}^a}{(d-1) A'}.
}
Using \eqref{eq:Einsteinfirstorder}, one can show that
\SP{
	\partial_z \mathfrak{b} = \Big[ - \frac{2d \Phi'_a}{d-1} + \frac{2 V_a }{(d-1) A'} + \frac{4 \Phi'_b \Phi'^b \Phi'_a}{(d-1)^2 A'^2} + \frac{2 \Phi'_a D_z}{(d-1) A'} \Big] \mathfrak{a}^a.
}
Plugging everything into \eqref{eq:scalarfirstorder2} finally gives us
\SP{
\label{eq:diffeq}
	&\Big[ D_z^2 + d A' D_z + e^{-2A} \Box \Big] \mathfrak{a}^a - \\ &\Big[ V^a_{|c} - \mathcal{R}^a_{\ bcd} \Phi'^b \Phi'^d + \frac{4 (\Phi'^a V_c + V^a \Phi'_c )}{(d-1) A'} + \frac{16 V \Phi'^a \Phi'_c}{(d-1)^2 A'^2} \Big] \mathfrak{a}^c = 0.
}

\section{Superpotential from the BPS Equations}

For Type A, it is possible to find a superpotential $W$ using the BPS equations. We will consider spinors in Type IIB SUSY variations that satisfy
\SP{
\epsilon=i\epsilon^*,\;\;\ 
\Gamma_{\theta\varphi}\epsilon=\Gamma_{\tilde{\theta}\tilde{\varphi}}\epsilon,\;\;\; 
\Gamma_{r\tilde{\theta}\tilde{\varphi}\psi}\epsilon= \epsilon.
} 
The gravitino 
variation $\delta \psi_x = 0$ gives

\SP{
	f' = \frac{e^{-2 f-2 g-2 h+\frac{\phi }{2}}}{16} \left(e^{2 g} \
N_f-\left(e^{2 g}-4 e^{2 h}\right) N_c\right)
}
where prime denotes differentiation with respect to $\rho$.
These equations are the same as the ones coming from the dilatino 
variations with $\phi = 4f$. 
Further, $\delta \psi_\theta = 0$ gives
\SP{
	h' = \frac{1}{4} e^{-2 (f+h)} \left(e^{\phi /2} N_c+e^{2 (f+k)}-e^{\phi \
/2} N_f\right),
}
while $\delta \psi_1 = 0$ gives
\SP{
	\tilde g' = e^{-2 \left(f+\tilde{g}\right)} \left(e^{2 (f+k)}-e^{\phi /2} N_c\right),
}
and, finally, $\delta \psi_3 = 0$ gives
\SP{
	k' = \frac{1}{4} e^{-2 \left( f+h+\tilde{g}\right)} \Big[& -e^{\phi /2} \
\left(4 e^{2 h}-e^{2 \tilde{g}}\right) N_c-4 e^{2 (f+h+k)}+ \\& 8 e^{2 \
\left(f+h+\tilde{g}\right)}-e^{2 \left(f+k+\tilde{g}\right)}-e^{\frac{\phi }{2}+2 \tilde{g}} N_f\Big].
}
The equation of motion for $A$ gives us an expression for the superpotential
\EQ{
	W = - \frac{3}{2} \frac{\partial A}{\partial z} = -\frac{e^{-A-k}}{2} \left( 8f' +2g' +2h' +k' \right),
}
where (as above) prime denotes differentiation with respect to $\rho$.
Using the above BPS equations, one arrives at
\SP{
	W = \frac{1}{16} e^{-g-2 (p+x)} \Big[ e^{6 p+x+\frac{\phi }{2}}
   \left(\left(-1+e^{2 g}\right) N_c+N_f\right)- \\ 4
   \left(1+e^{2 g}+2 e^{g+6 p+2 x}\right)\Big].
}
One can check that this superpotential $W$ reproduces the Type A 5d potential $V$ through
\EQ{
	V = \frac{1}{2} G^{ab} W_a W_b - \frac{4}{3} W^2.
}
Also, the equations of motion derived from the superpotential are precisely the BPS equations given above.

\section{Numerical Methods}

Suppose that we have a system of $n$ scalar fields satisfying a second order linear differential equation, and that the boundary conditions in the IR single out $p$ linearly independent solutions, whereas the boundary conditions in the UV single out $q$ solutions. A solution is completely characterized by evaluating it and and its derivative at a chosen point. Therefore, let us form vectors $({\mathfrak{a}_{IR}}_{(i)}, \partial_\rho {\mathfrak{a}_{IR}}_{(i)})$ where different $i$ denote different solutions in the IR, and we have suppressed the field index. These are $p$ column vectors of size $2n$. By evolving them numerically from the IR, we can evaluate them at any point we like, and therefore they are functions of $\rho$. Similarly, we form $q$ column vectors from the UV solutions, $({\mathfrak{a}_{UV}}_{(i)}, \partial_\rho {\mathfrak{a}_{UV}}_{(i)})$. The question that we need to answer is whether, for a particular value of $K^2$, we can find a solution that interpolates between the correct IR and UV behaviours. In other words, we want to know whether we can find a linear combination of the $p$ solutions in the IR and write it in terms of a linear combination of the $q$ solutions in the UV. For $p = q = n$, this is true if and only if the deterimant of the matrix formed by putting the IR and UV column vectors next to each other is equal to zero. It is convenient to evaluate this matrix at a point chosen between the IR and UV. In other words, the linearly independent solutions satisfying the boundary conditions in the IR and in the UV, respectively, are evolved numerically to a midpoint, where the determinant is evaluated. If it is zero for a particular value of $K^2$, there is a pole in the correlator. This is the midpoint determinant method described in \cite{Berg:2006xy}.

We would now like to generalize this method to include cases where $p$ and $q$ are not necessarily equal to $n$. In such cases, the matrix obtained by putting the IR and UV column vectors next to each other is not generally a square matrix, and therefore we can not answer the question of whether the vectors are linearly independent by evaluating a determinant. The method we will use instead is the following. First we normalize the vectors $({\mathfrak{a}_{IR}}_{(i)}, \partial_\rho {\mathfrak{a}_{IR}}_{(i)})$ and $({\mathfrak{a}_{UV}}_{(i)}, \partial_\rho {\mathfrak{a}_{UV}}_{(i)})$. Let us denote by $X^a_{\ i}$ ($i = 1, \ldots, p + q$, $a = 1, \ldots, 2n$) the matrix formed by putting these normalized column vectors next to each other. Then we construct an orthonormal basis $e^i_{\ a}$ ($i = 1, \ldots, p + q$, $a = 1, \ldots, 2n$) for the subspace spanned by these vectors. Finally, we project the normalized vectors onto the basis and form a matrix $Y^i_{\ j} = e^i_{\ a} X^a_{\ j}$. This is now the $(p+q) \times (p+q)$ matrix whose determinant we compute at a midpoint between the IR and UV. Again, if it is equal to zero, there is a pole in the correlator.

\end{document}